\documentclass[11pt]{article}
\usepackage[utf8]{inputenc}
\usepackage[T1]{fontenc}
\usepackage[width=18cm,height=24cm]{geometry}
\usepackage{amsmath,amsfonts,amssymb}
\usepackage{indentfirst}
\usepackage{hyperref}
\usepackage{graphicx}
\usepackage{xcolor}
\usepackage{enumitem}
\usepackage{mathtools}
\usepackage{dsfont}
\usepackage{tikz}
\usetikzlibrary{decorations.markings}
\usetikzlibrary{calc}

\DeclareMathOperator{\Tr}{Tr}
\DeclareMathOperator{\tr}{tr}
\DeclareMathOperator{\Ad}{Ad}
\DeclareMathOperator{\ad}{ad}
\DeclareMathOperator{\diag}{diag}
\DeclareMathOperator{\pexp}{Pexp}
\newcommand{\calR}{\mathcal{R}}
\newcommand{\calA}{\mathcal{A}}
\newcommand{\calO}{\mathcal{O}}
\newcommand{\id}{I}

\numberwithin{equation}{section}

\newcommand\ooval{}
\def\ooval(#1,#2,#3){%
	\begin{scope}[thick]
		\draw (#1,#2) -- (#1+3.,#2);
		\draw (#1,#2+1.2) -- (#1+3.,#2+1.2);
		\draw (#1,#2+1.2) arc (90:270:0.6);
		\draw (#1+3.,#2) arc (-90:90:0.6);
		\draw (#1+1.5, #2+0.27) node[anchor=south]{#3};
	\end{scope}
}

\hypersetup{
	colorlinks,%
	citecolor=purple,%
	filecolor=purple,%
	linkcolor=purple,%
	urlcolor=purple
}

\title{\bf Large $k$ topological quantum computer}
\date{}

\author{Nikita Kolganov$^{a,b,c}$\thanks{\href{mailto:nikita.kolganov@phystech.edu}{nikita.kolganov@phystech.edu}}, Sergey Mironov$^{d,c,b,a}$\thanks{\href{mailto:sa.mironov_1@physics.msu.ru}{
sa.mironov\_1@physics.msu.ru}}, and Andrey Morozov$^{e,c,a}$\thanks{\href{mailto:andrey.morozov@itep.ru}{andrey.morozov@itep.ru}}}

\begin{document}
	\vspace{-0.5cm}
	\begin{flushright}
		ITEP-TH-11/21 \\
		IITP-TH-8/21\\
		MIPT-TH-7/21
	\end{flushright}

\vspace{-1cm}
	
{\let\newpage\relax\maketitle}

\vspace{-1cm}

\begin{center}
    $^a$ {\small {\it Moscow Institute of Physics and Technology, 141701 Dolgoprudny, Russia}}\\
    $^b$ {\small {\it Institute for Theoretical and Mathematical Physics, MSU, 119991 Moscow, Russia}}\\
	$^c$ {\small {\it Institute for Theoretical and Experimental Physics, 117218 Moscow, Russia}}\\
	$^d$ {\small {\it Institute for Nuclear Research, RAS, 117312 Moscow, Russia}} \\
	$^e$ {\small {\it Institute for Information Transmission Problems, 127994 Moscow, Russia}}\\
\end{center}

\begin{abstract}
    Chern-Simons topological quantum computer is a device that can be effectively described by the Chern-Simons topological quantum field theory and used for quantum computations. Quantum qudit gates of this quantum computer are represented by sequences of quantum $\calR$-matrices. Its dimension and explicit form depend on the parameters of the Chern-Simons theory --- level $k$, gauge group $SU(N)$, and representation, which is chosen to be symmetric representation $[r]$. In this paper, we examine the universality of such a quantum computer. We prove that for sufficiently large $k$ it is universal, and the minimum allowed value of $k$ depends on the remaining parameters $r$ and $N$.
\end{abstract}

\section{Introduction}
   Quantum computers potentially provide very high performance in quantum systems simulation and some cryptography problems. Unfortunately, quantum computers that exist and are being developed suffer from errors, because they are open systems and unavoidably interact with an environment. These errors interfere with effective quantum computations. Now, several potential solutions to the quantum errors problem are known. Some of them can be grouped under the name of quantum error-correcting codes and mostly based on the redundancy of the system \cite{kitaev}. The other idea is based on the fact that quantum errors are local, and the states, which can be represented as some topological objects like braids, are not subject to such errors. Quantum computers that exploit this fact are called topological quantum computers (see \cite{topo-review} and references therein).
   
   While some topological quantum computers are based on the construction of topological states in the usual quantum computers, the others use the systems in which all the states are topological as is. One of such models is Chern-Simons (2+1)-dimensional topological quantum field theory, whose observables depend only on the topology of space-time trajectories of charged particles. More precisely, these observables are equal to knot (or link) invariants corresponding to such trajectories \cite{witten}.
   
   Topological quantum computation based on Chern-Simons theory \cite{kitaev-freedman,towards} has some difficulties. First of all, the naive Hilbert space of the theory, which is the tensor product of particles representation, is very redundant. Indeed, elementary operations (e.g. particle interchange) act only on the invariant subspace of the naive Hilbert space, the space of conformal blocks, or so-called Verlinde space. Therefore, elementary operations in Chern-Simons quantum computer are not the sets of one-, two-, etc. gates but one-qudit gates, where the dimension of qudit is the dimension of the Verlinde space. At this point, technical difficulty arises, namely, quantum Racah matrices, responsible for explicit form braiding operations ($\calR$-matrices) and acting on irreducible subspace, are known only for some particular groups and their representations.
   
   At the same time, an effective quantum computer should be universal. This means that an arbitrary unitary matrix can be represented by a sequence of elementary operations with any required accuracy, which are $\calR$-matrices for the Chern-Simons quantum computer. Despite positive results for some very particular parameters of the Chern-Simons theory (mostly, $SU(2)$ case) \cite{freedman, simon}, in general, it is not known if this is possible.
   
   In this paper, we focus on the Chern-Simons topological quantum computer with the arbitrary level $k$, group $SU(N)$, and particles in arbitrary symmetric representation $[r]$ and address the question for which $k$, $N$, and $r$ such quantum computer is universal. This case is rather general and important for the following reasons. First of all, Hilbert space has the dimension $r+1$ so that $n = \lfloor\log_2 (r+1)\rfloor$ qubit quantum computer can be implemented. Moreover, expressions for $\calR$-matrices are known \cite{6j-orig,6j-racah}, and universality can be examined explicitly. For this purpose, we adopt recent general results on the universality of one-qudit gates \cite{qudit, qudit_short} and find the range of parameters $k$, $N$, and $r$ that lead to the universal quantum computer. In short, universality takes place for sufficiently large values of $k$ and it's minimal allowed value depends on $r$ and $N$. For the fixed $N$ it behaves as $O(r^{5/2})$. This extends previous result \cite{Kolganov} for $r=1$ case.
   
   The rest of the paper is organized as follows. In Sect. \ref{CS_sec} we review the Chern-Simons theory and computational tools for knot invariants, which are important in what follows. In Sect. \ref{QC_sec} we describe the Chern-Simons quantum computer, i.e. how the quantum computations can be performed with a device described by the Chern-Simons theory. In Sect. \ref{Univ_sec} we derive a universality condition and discuss in Sect. \ref{Disc_sec}, why the obtained restrictions on the Chern-Simons theory parameters look natural.

\section{Chern-Simons theory and knot invariants} \label{CS_sec}

    The Chern-Simons theory is a three-dimensional topological quantum field theory, defined by the following action
    \begin{equation}
        S_\mathrm{CS}[\calA] = \frac{k}{4\pi} \int_{S^3} \Tr \left[ \calA \wedge d\calA + \tfrac23 \calA \wedge \calA \wedge \calA\right],
    \end{equation}
    where $\calA = \calA_\mu d x^\mu$ is gauge connection of $SU(N)$ group, which transforms in adjoint representation.
    Gauge-invariant observables of such theory are averages of Wilson loops
    \begin{equation}
        \langle W_V^K \rangle_\mathrm{CS} = \left\langle \Tr_R \pexp \oint_K \calA \right\rangle_\mathrm{CS}, \label{w_loop}
    \end{equation}
    where $K$ is a closed contour of integration, $V$ is representation in which the connection $\calA$ is taken. These averages are equal to knot invariants --- HOMFLY-PT polynomials \cite{homfly,morozov-homfly}, whose arguments $q$, $A$ are defined by the parameters of the theory
    \begin{equation}
        q = \exp\frac{2\pi i}{N + k}, \qquad A = q^N.
    \end{equation}

	Close connection of the $SU(N)$ Chern-Simons theory with $SU(N)_k$ WZNW conformal field theory allows to identify it's Hilbert space with the space conformal blocks --- Verlinde space \cite{witten,conf_bl}. Evolution operators on such space are braiding operators, which performs the half-monodromies of conformal block arguments. These are called $\calR$-matrices and satisfy the Yang-Baxter equation	
    \begin{equation}
        \calR_i \, \calR_{i+1} \, \calR_i = \calR_{i+1} \, \calR_i \, \calR_{i+1},
    \end{equation}
    where $i$ denotes the position of conformal block arguments to be exchanged.
    
    Most simple nontrivial situation of four-point conformal blocks, corresponding to the case in which the knot $K$ can be represented as four-plat knot, i.e. for any time slice is punctured by Wilson line four times. In this case the Hilbert space is isomorphic to invariant subspace of tensor product of two representations $V$ and two conjugate representations $\bar V$, namely
    \begin{equation}
        \mathcal H = \operatorname{inv} (V \otimes \bar V \otimes V \otimes \bar V). \label{Hilb_sp}
    \end{equation}

	\begin{figure}
		\centering
		\begin{tikzpicture}
			\begin{scope}[thick]
				\draw (0., 1.2) node[anchor=south]{$V_1$}  -- (0.5,0.6) -- (0.5,0) -- (2.5,0) -- (2.5,0.6) -- (3., 1.2) node[anchor=south]{$V_4$};
				\draw (1., 1.2) node[anchor=south]{$V_2$} -- (0.5,0.6);
				\draw (2.5,0.6) -- (2., 1.2) node[anchor=south]{$V_3$};
				\draw (1.5, 0.) node[anchor=south]{$t$};
			\end{scope}
		\end{tikzpicture} \qquad
		\begin{tikzpicture}
			\begin{scope}[thick]
				\draw (0., 1.2) node[anchor=south]{$V_1$}  -- (0.,0.) -- (3.,0.) -- (3.,0) -- (3., 1.2) node[anchor=south]{$V_4$};
				\draw (1., 1.2) node[anchor=south]{$V_2$} -- (1.5,0.6) -- (2., 1.2) node[anchor=south]{$V_3$};
				\draw (1.5, 0.6) -- (1.5,0.);
				\draw (1.5, 0.3) node[anchor=west]{$s$};
			\end{scope}
		\end{tikzpicture}
		\caption{Two bases in the invariant subspace of $V_1 \otimes V_2 \otimes V_3 \otimes V_4$ correspond to two different fusions of representation.}
		\label{fus_fig}
	\end{figure}
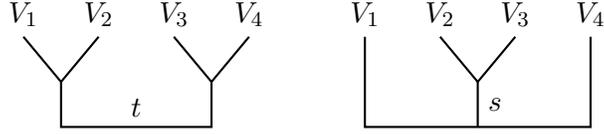
	It has two bases, so-called $t$- and $s$-basis, corresponding to two different fusions of representations (see Fig. \ref{fus_fig}). Here $t$ and $s$ denotes the intersections of the fused representations, namely $t\in (V_1 \otimes V_2) \cap (\bar V_3 \otimes \bar V_4)$, $s\in (V_2 \otimes V_3) \cap (\bar V_1 \otimes \bar V_4)$. Matrices $\calR_1$ and $\calR_3$, performing the half-monodromies of the first and third representations (or the arguments of the corresponding conformal blocks), are diagonal in $t$-basis, while $\calR_2$ that performs the half-monodromy of the second pair of representations is diagonal in $s$-basis. The transition matrices between two bases are called $q$-Racah coefficients, and depend on the representations fused. In what follows we will work in $t$-basis. Eigenvalues of $\calR$-matrix, acting on the pair of the representations $V_1$, $V_2$, are given by the formula
	\begin{equation}
		\lambda_{V_1,V_2,t} = \epsilon_{V_1,V_2,t} \, q^{\frac12(Q_1 + Q_2 - Q_t)}, \qquad \epsilon_{V_1,V_2,t} = \pm 1,
	\end{equation}
	where $Q_{1,2}$, $Q_t$ are the quadratic Casimirs of the corresponding representations. Thus, we fully describe the braiding operations that define the evolution of the Chern-Simons Hilbert space. It remains to determine one left operation, which is contraction of the Wilson lines --- it is defined by the projection on the trivial representation $t = \varnothing$.
	
\begin{figure}[h!]
        \centering
        \includegraphics[angle=270]{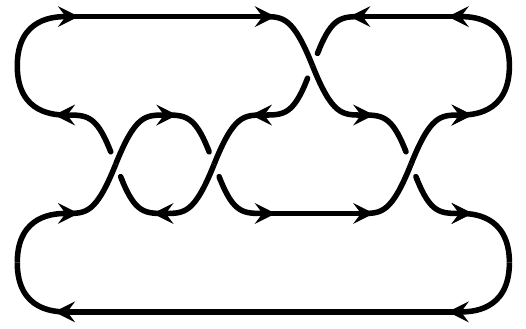}
        \caption{The figure-eight knot $4_1$ in four-plat form. The corresponding braid is $\calR_2^{-1} \calR_1^{\phantom{-1}} \!\!\! \calR_2^{-1} \calR_2^{-1}$.}
        \label{f_eight}
    \end{figure}

	Thus, to calculate the HOMFLY-PT polynomial of a knot in the representation $V$, one first should represent it as four-plat knot (if possible) and assign the direction to the strand (see Fig. \ref{f_eight}). Now, one interprets forward directed strands to the representation $V$, whereas backward directed strands correspond to its conjugate representation $\bar V$. Strand crossing represents the action of $\calR$-matrix on the corresponding pair of representations and caps at the beginning and at the end of the braid correspond to the projection on the trivial representation $\varnothing$. Therefore, to compute Wilson loop (\ref{w_loop}), one should calculate a product of $\calR$-matrices, corresponding to the knot's picture (projection), and take it's matrix element, corresponding to the trivial representation $\varnothing$. See Fig. \ref{f_eight} for instructive example.

	Despite the fact that the calculation of $q$-Racah coefficients of the $SU(N)$ group is a very difficult and generally unsolved problem, these are known for the symmetric representations and its conjugates \cite{6j-orig,6j-racah,Alekseev}, i.e. $V$ is defined by the one-row Young diagram $[r]$, where $r$ is length of the row. For this particular choice, quantum Racah coefficients are given by
    \begin{align}
        \bar S &= \epsilon_{\{V_i\}} \sqrt{\dim_q V_{12} \dim_q V_{23}} \cdot
        \begin{Bmatrix}
            V_1 & \bar{V}_2 & V_{12} \\
            V_3 & \bar{V}_4 & V_{23}
        \end{Bmatrix}, \label{rac_1st}\\
        S &= \epsilon_{\{V_i\}} \sqrt{\dim_{q} V_{12} \dim_{q} V_{23}} \cdot \begin{Bmatrix}
            V_{1} & V_{2} & V_{12} \\
            \bar{V}_{3} & \bar{V}_{4} & V_{23}
        \end{Bmatrix}. \label{rac_2nd}
    \end{align}
    Here braces denote quantum $6j$-symbols, $\dim_q V$ is quantum dimension of representation $V$, and $\epsilon_{\{V_i\}} = \pm 1$ and given explicitly below.
    In the case of $\bar S$ corresponding representations are $V_1=V_3 = [r]$, $\bar V_2 = \bar V_4 = [\bar r]$, while $V_{12}$ and $V_{23}$ are terms of tensor product decomposition
    \begin{equation}
        [r] \otimes [\bar r] = \bigoplus_{n=0}^r [2n, n^{N-2}], \label{r_bar_r}
    \end{equation}
    where $[2n, n^{N-2}]$ denotes $\mathfrak{su}_N$ representation defined by this Young diagram. Corresponding quantum dimension of $[2n, n^{N-2}]$ can be calculated by the use of general formula and turns out to be
    \begin{equation}
        \dim_q [2n, n^{N-2}] = \frac{[N+2n-1] [N+n-2]!^2}{[n]!^2 [N-1]! [N-2]!},
        \label{q_dim_rrb}
    \end{equation}
    where $[m] = (q^m-q^{-m})/(q-q^{-1})$ denotes a quantum number\footnote{We use the same square bracket notation for Young diagrams and the quantum numbers since it arise in different context and will not lead to any confusion.}.
    Denoting $V_{12}$ and $V_{34}$ as $i$ and $j$, correspondingly, quantum 6$j$-symbols, involved in (\ref{rac_1st}) can be explicitly written as
    \begin{equation}
        \begin{Bmatrix}
            r & \bar r & i \\
            r & \bar r & j
        \end{Bmatrix} = 
        \frac{[i] !^{2}[j] !^{2}[r-i] ![r-j] ![N-1] ![N-2] !}{[r+i+N-1] ![r+j+N-1] !} \sum_{z}(-)^{z} \frac{[r+N-1+z] !}{[z-i] !^{2}[z-j] !^{2}[r-z] ![i+j-z] ![i+j+N-2-z] !}
    \end{equation}
	For brevity, we denote the representations $[r]$, $[\bar r]$ simply by $r$, $\bar r$, respectively. Specifying $\epsilon_{\{R_i\}}$ as $(-)^{i+j+r}$, we find that all quantities in (\ref{rac_1st}) become explicitly defined.
     \begin{figure}[h!]
    	\centering
    	\begin{tikzpicture}
    		\ooval(0,0,$\boldsymbol{[\bar r]\otimes[ \bar r]\otimes[r]\otimes[r]}$)
    		\ooval(0,2,$\boldsymbol{[\bar r]\otimes[r]\otimes[ \bar r]\otimes[r]}$)
    		\ooval(3.5,4,$\boldsymbol{[\bar r]\otimes[r]\otimes[r]\otimes[\bar r]}$)
    		\ooval(-3.5,4,$\boldsymbol{[r]\otimes[\bar r]\otimes[ \bar r]\otimes[r]}$) 
    		\ooval(0,6,$\boldsymbol{[r]\otimes[\bar r]\otimes[r]\otimes[\bar r]}$)  
    		\ooval(0,8,$\boldsymbol{[r]\otimes[r]\otimes[\bar r]\otimes[\bar r]}$)
    		\draw[thin,dashed,gray] (6,1.6) -- (-3,7.6);
    		\begin{scope}[thick]
    			\draw [stealth-](1.1, 1.3) -- (1.1, 1.9);
    			\draw [-stealth](1.9, 1.3) -- (1.9, 1.9);
    			\draw (1.9, 1.6) node[anchor=west]{$\mathcal R_2 =\bar S \bar T S$};
    			\draw (1.1, 1.6) node[anchor=east]{$\mathcal R_2 = S^T \bar T \bar S$};
    			\draw [stealth-](1.1, 7.3) -- (1.1, 7.9);
    			\draw [-stealth](1.9, 7.3) -- (1.9, 7.9);
    			\draw (1.9, 7.6) node[anchor=west]{$\mathcal R_2 = S^T \bar T \bar S$};
    			\draw (1.1, 7.6) node[anchor=east]{$\mathcal R_2 = \bar S \bar T S$};
    			\draw [stealth-](1.9,-0.1) arc (0:-180:0.4);
    			\draw (1.5, -0.5) node[anchor=north]{$\mathcal R_1 = \mathcal R_3 = T$};
    			\draw [stealth-](1.9, 9.3) arc (0:180:0.4);
    			\draw (1.5, 9.7) node[anchor=south]{$\mathcal R_1 = \mathcal R_3 = T$};
    			\draw [-stealth](3.4,3.3) -- (3.8, 3.9);
    			\draw (3.6, 3.5) node[anchor=west]{$\mathcal R_3 = \bar T$};
    			\draw [stealth-](2.8,3.3) -- (3.2, 3.9);
    			\draw (3.1, 3.7) node[anchor=east]{$\mathcal R_3 = \bar T$};
    			\draw [-stealth](-0.4,3.3) -- (-0.8, 3.9);
    			\draw (-0.6, 3.5) node[anchor=east]{$\mathcal R_3 = \bar T$};
    			\draw [stealth-](0.2,3.3) -- (-0.2, 3.9);
    			\draw (-.1, 3.7) node[anchor=west]{$\mathcal R_3 = \bar T$};
    			\draw [stealth-](3.8,5.3) -- (3.4, 5.9);
    			\draw (3.6, 5.7) node[anchor=west]{$\mathcal R_3 = \bar T$};
    			\draw [-stealth](3.2,5.3) -- (2.8, 5.9);
    			\draw (3.1, 5.5) node[anchor=east]{$\mathcal R_3 = \bar T$};
    			\draw [stealth-](-0.8,5.3) -- (-0.4, 5.9);
    			\draw (-0.6, 5.7) node[anchor=east]{$\mathcal R_3 = \bar T$};
    			\draw [-stealth](-0.2,5.3) -- (0.2, 5.9);
    			\draw (-.1, 5.5) node[anchor=west]{$\mathcal R_3 = \bar T$};
    			\draw [stealth-](7.2, 4.4) arc (-150: 150:0.4);
    			\draw (8., 5.) node[anchor=south]{$\mathcal R_2 = S T S^T$};
    			\draw [stealth-](-4.2, 4.4) arc (-30: -330:0.4);
    			\draw (-5., 5.) node[anchor=south]{$\mathcal R_2 = S T S^T$};
    		\end{scope}

    	\end{tikzpicture}
    	\caption{Graph, representing the action of $\calR$-matrices in corresponding representations. Any four-plat knot can be deformed such that one of the side strands becomes unused, and only braiding operations above (or below) the dashed line are employed.}
    	\label{repr_graph}
    \end{figure}
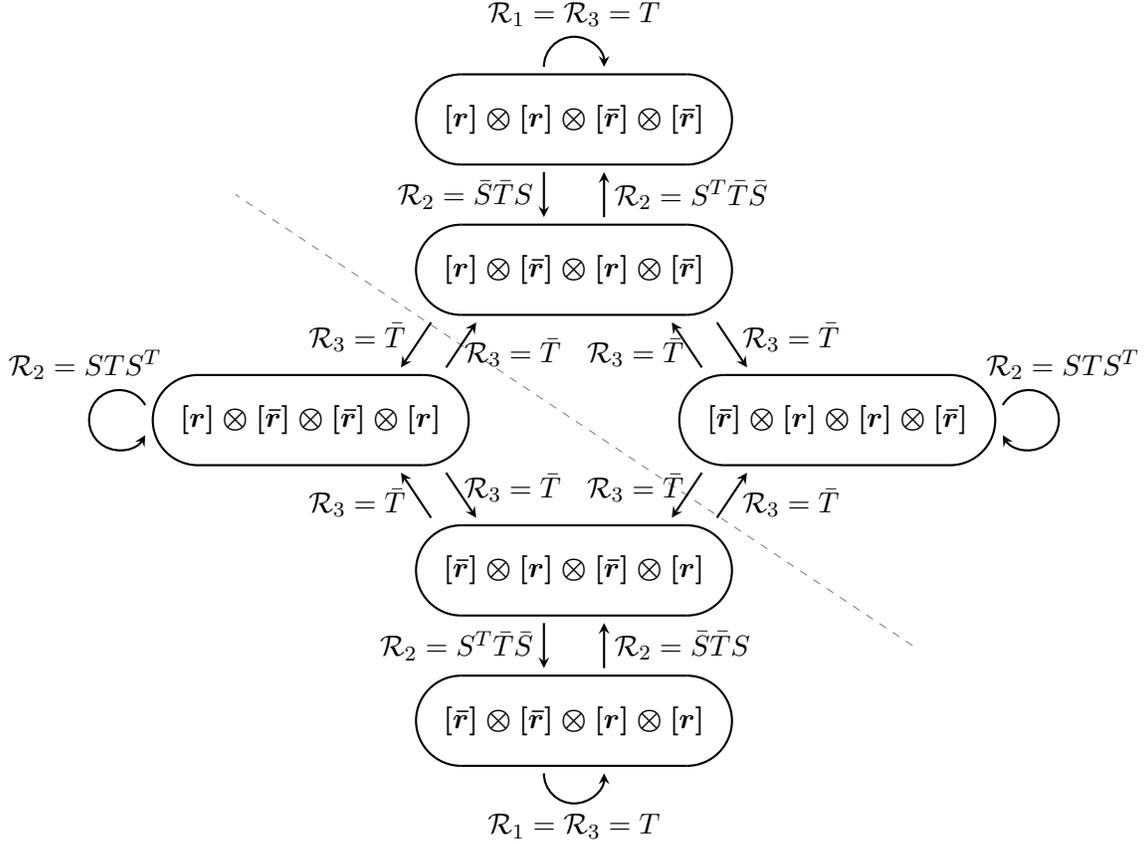
    
    Similarly, let us clarify the formula (\ref{rac_2nd}). Here $V_1 = V_2 = [r]$, $\bar V_3 = \bar V_4 = [\bar r]$, $V_{23}$ are irreducible representations in the decomposition (\ref{r_bar_r}) of the product $[r]\otimes[\bar r]$, while $V_{12}$ are irreducible representation if the product
    \begin{equation}
        [r] \otimes [r] = \bigoplus_{n=0}^r [r+n, r-n]. \label{r_r}
    \end{equation}
    Denoting $V_{12}$ and $V_{23}$ as $i$ and $j$, correspondingly, we can write the explicit form of quantum 6$j$-symbols involved in (\ref{rac_2nd}) as
    \begin{equation}
        \begin{Bmatrix}
            r &  r & i \\
            \bar r & \bar r & j
        \end{Bmatrix} = 
        \frac{[i] !^{2}[j] !^{2}[r-i] ![r-j] ![N-1] ![N-2] !}{[r+i+N-1] ![r+j+N-1] !} \sum_{z}(-)^{z} \frac{[r+N-1+z] !}{[z-i] !^{2}[z-j]!^{2}[r-z] ![i+j-z] ![z-j+N-2] !}
    \end{equation}
    Quantum dimensions of the decomposition (\ref{r_r}) components are
    \begin{equation}
        \dim_q [r+n, r-n] = \frac{[2n+1] [N+r+n-1]! [N+r-n-2]!}{[N-1]! [N-2]! [r-n]! [r+n+1]!},
        \label{q_dim_rr}
    \end{equation}
    and $\epsilon_{\{R_i\}} = (-)^{i+j+r}$ again.
    
    Eigenvalues of the $\calR$-matrices are given by the diagonal matrices
    \begin{align}
    	T &= \diag\left((-1)^{m+1} q^{-r^{2}+m^{2}+m}A^{-r}\right), & m&=0,\ldots r, \label{T_matr}\\
    	\bar T &= \diag\left( (- q^{m-1} A)^m \right), & m&=0,\ldots r, \label{Tbar_matr}
    \end{align}
	where $T$ corresponds to the action on the pair of equal representations, while $\bar T$ --- on the pair opposite representations. Since we work in $t$-basis, $\calR_1$, $\calR_3$ are given by $T$ or $\bar T$, while $\calR_2$ is given by $T$ or $\bar T$ conjugated by $S$ and/or $\bar S$. The types of $\calR$-matrices in this basis, depending on the order of the representations, are depicted at Fig. \ref{repr_graph}.
       
\section{Chern-Simons topological quantum computer} \label{QC_sec}
	
	One of the most promising application of a quantum computers is a simulation of quantum systems. Usually, a Hilbert space of the system to be simulated is approximated by the finite-dimensional Hilbert space of a quantum computer, and then a unitary evolution of the quantum system is approximated by a sequence of unitary quantum gates. Subsequent measurements of the state allow to explore the physical properties of the simulated system.
	
	Thus, to perform quantum computations of such type (which are usually called universal quantum computations), a quantum computer should have sufficiently large Hilbert space and be able to do the following operations
	\begin{enumerate}[parsep=0pt]
		\item preparation of an arbitrary state,
		\item application of an arbitrary unitary transformation to the state,
		\item performing projective measurements in an arbitrary basis.
	\end{enumerate}
	Obviously, if the second condition is satisfied, the first and the third can be reduced to the single state preparation and measurements in a single basis, respectively.
	
	Now, suppose we have a device, which can be described by Chern-Simons theory. How to perform quantum computations using this device? To answer this question, we should define what are exactly state preparation, unitary evolution, and the measurement process in the Chern-Simons theory.
	Wilson loop average $\langle W_V^K\rangle$ can be thought of as an amplitude of propagation of the particle, charged in the representation $V$, through the path $K$. For given space-time foliation it can be recast as particle-antiparticle creation, braiding of its worldlines, and subsequent annihilation into vacuum.	
	
	In the case of four-plat knot trajectories $K$, foliation is obvious, so that initial state preparation correspond to the creation of two particle-antiparticle pairs in the representations $V$, $\bar V$ in the trivial overall representation.
	The Hilbert space (\ref{Hilb_sp}), in which this state lives, is invariant subspace of the tensor product $V \otimes \bar V \otimes V \otimes \bar V$. In the case $V=[r]$ of our interest, it's dimension equals to the number of terms on the r.h.s of (\ref{r_bar_r}), which is $r + 1$, so we need to sufficiently large $r$ for the simulation to be effective. In the usual quantum computing terms, we have one-qudit with $d = r + 1$ or, equivalently, $n = \lfloor \log_2 (r+1) \rfloor$ qubits.
	Thus, in the $t$-basis, defined in Sect. \ref{CS_sec}, creation of two particle-antiparticle pairs charged by the symmetric representation $[r]$ corresponds to preparation of the vector $|\varnothing\rangle = (1, 0, \ldots 0)$.
	
	Evolution is given by braiding operations --- action of $\calR$-matrices, whose explicit form was introduced in Sect. \ref{CS_sec}. These are not unitary by default, so we should deduce a condition, under which the unitarity of $\calR$-matrices does holds.
	These are composed of matrices $T$, $\bar T$, $S$, $\bar S$ and its inverses, defined in (\ref{rac_1st})--(\ref{rac_2nd}), (\ref{T_matr})--(\ref{Tbar_matr}).
	The matrices $T$, $\bar T$ are always unitary, since these are diagonal and its diagonal elements are roots of unity due to $|q|=|A|=1$.
	At the same time, the matrices $S$, $\bar S$ are orthogonal, but not necessarily real-valued, since they involve square root of $q$-numbers that can be negative. The maximal $q$-number involved is $[N+2r-1] = \sin(2\pi\frac{N+2r-1}{N+k}) / \sin(\frac{2\pi}{N+k})$, and sufficient positivity condition is
	\begin{equation}
		k > N + 4r - 2,
	\end{equation}
	under which $S$, $\bar S$ are orthogonal and real, hence it follows that these are unitary. Thus, all braiding operations, defining an evolution, are unitary under the condition above.	
	
	Finally, measurement corresponds to annihilation of the pairs into vacuum and projection onto trivial representation that in $t$-basis corresponds to contraction with a ket-vector $\langle\varnothing| = (1, 0, \ldots 0)$. The result obtained equals to amplitude of particle propagation through the contour of knot, defined by contracted braid, on the one hand, and to the HOMFLY-PT polynomial of this knot, on the other hand.

	Let us demonstrate the procedure on the example of the figure-eight knot, depicted at Fig. \ref{f_eight}, and introduce intuitive graphical notation. We first prepare two particle-antiparticle pairs, corresponding tho the state
	\begin{equation}
		|\varnothing\rangle = (\underbrace{1, 0, \ldots 0}_{r+1})^\dagger \; = \; \parbox[c]{3.5cm}{\includegraphics{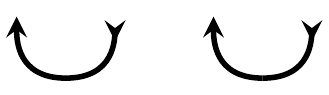}}.
	\end{equation}
	Next, we appropriately braid their worldlines that results into the action of corresponding $\calR$-matrices 
	\begin{equation}
		\calR_2^{-1} \calR_1^{\phantom{-1}} \!\!\! \calR_2^{-1} \calR_2^{-1} \quad=\quad \parbox[c]{4.4cm}{\includegraphics{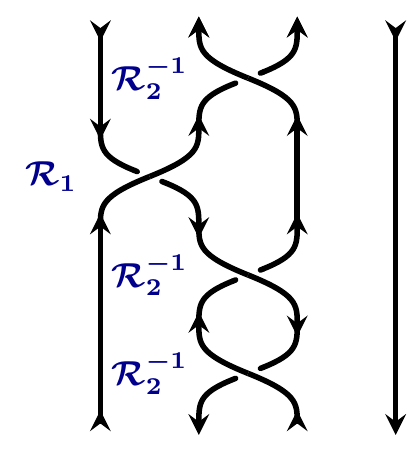}}
	\end{equation}
	on the prepared vector state. Measurement process is defined by annihilation of the pairs into vacuum and projection onto trivial representation
	\begin{equation}
		\langle\varnothing| = (\underbrace{1, 0, \ldots 0}_{r+1}) \; = \; \parbox[c]{3.5cm}{\includegraphics{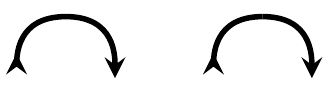}}.
	\end{equation}
	The whole process of particles propagation and it's amplitude that equals to the Wilson loop average, can be represented as
	\begin{equation}
			\langle W_V^K \rangle_\mathrm{CS} = \langle \varnothing |\calR_2^{-1} \calR_1^{\phantom{-1}} \!\!\! \calR_2^{-1} \calR_2^{-1} |\varnothing\rangle  =
			(\calR_2^{-1} \calR_1^{\phantom{-1}} \!\!\! \calR_2^{-1} \calR_2^{-1})_{11}
			\; = \;\;
		\parbox[c]{4.4cm}{\includegraphics{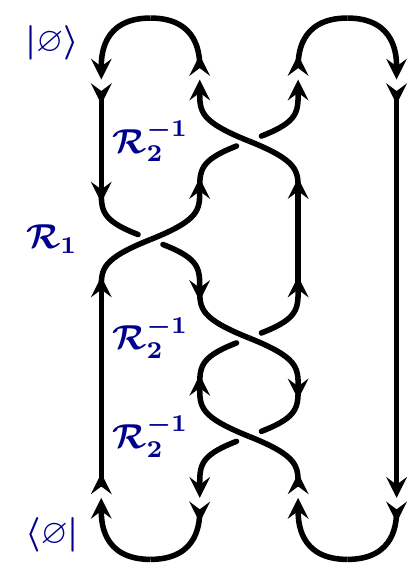}}.
	\end{equation}

	Thus, having the Chern-Simons topological quantum computer, we are able to prepare a single state, act by the $\calR$-matrices (which are unitary under the condition (\ref{q_nonv})) on it, and measure projection onto single state. This quantum computer is universal, if arbitrary unitary matrix of appropriate dimension can be decomposed into a braiding operation --- a product of $\calR$-matrices. In the next section, we derive the condition under which this is indeed the case.

\section{Quantum computations at large $k$} \label{Univ_sec}

    In the previous section we discussed a physical realization of Chern-Simons topological quantum computer and found that to make it work efficiently, one should be able to prepare arbitrary initial states, represent arbitrary unitary transformation matrix via some product of $\calR$-matrices, and measure projections onto arbitrary states. The first and the third problems reduce to the second one.
    
    A set of unitary matrices is called a universal set of gates if an arbitrary unitary matrix can be represented as a product of matrices from a set of finite length with any predetermined accuracy. Thus, we need to show that the combinations of $\calR$-matrices, composed of $T$, $\bar T$, $S$, $\bar S$ and its inverses, being multiplied according to Fig. \ref{repr_graph} constitute a universal set of gates. As we will show below, some convenient subset of such matrices is universal, i.e. it generates a dense subset of unitary group, if the level $k$ of Chern-Simons theory satisfies some condition, which depends on the representation $[r]$ and group $SU(N)$.
    
    According to the theorems III.3 and IV.6 from \cite{qudit}, a set of gates $G = \{g_1, \ldots g_n\}$, $g_i \in SU(d)$ is universal, if $g_i$ belongs to the following vicinity of $SU(d)$ group center
    \begin{equation}
        \|g_i- e^{2i\pi k / d} \id \| < 1/\sqrt{2}, \quad Z(SU(d)) = \{e^{2i\pi k / d} \, \id, \; k = 0, \ldots (d-1) \}, \label{vicinity}
    \end{equation}
    while a set of Lie-algebra\footnote{Hereinafter we use anti-hermitean basis in the defining representation of $\mathfrak{su}_{d}$ algebra.} elements $X = \{x_1, \ldots x_n \}$, $x_i \in \mathfrak{su}_{\,d}$, such that $g_i = e^{x_i}$ satisfy a property
    \begin{equation}
        C(X) \coloneqq \{L \in \mathfrak{gl}_{d^2-1}: \; [L, \operatorname{ad}_{x_i}] = 0, \; i = 1, \ldots n\} = \{\lambda \,\id, \; \lambda  \in \mathbb R\}. \label{ncond}
    \end{equation}
	Here $\|\bullet\|$ defines a trace norm $\|A\| = \tr \sqrt{A A^\dagger}$. In other words, the conditions (\ref{vicinity}), (\ref{ncond}) state that the gates should be sufficiently close to the elements of center, and the set of its logarithms in adjoint representations should commute only with the identity matrix.
    
    As a set $G$ we choose the elements $\bar T^2$ and $\bar S \bar T^2 \bar S$, that is
    \begin{equation}
        G = \{\bar T^2, \; \bar S \bar T^2 \bar S\}, \label{rgates}
    \end{equation}
    where we will assume that $\bar T$ is multiplied by a phase factor, such that $\det \bar T^2 = 1$. According to Fig. \ref{repr_graph}, gates from this set map representation $[r]\otimes[\bar r]\otimes[r]\otimes[\bar r]$ on itself and, therefore, can be multiplied in arbitrary order. Graphically, these elements correspond to the following braiding operations
    \begin{align}
    	\bar T^2 &= \parbox[c]{2.5cm}{\includegraphics[width=2.5cm]{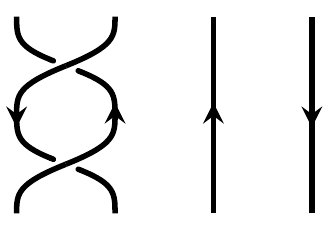}}, &
    	\bar S \bar T^2 \bar S &= \parbox[c]{2.5cm}{\includegraphics[width=2.5cm]{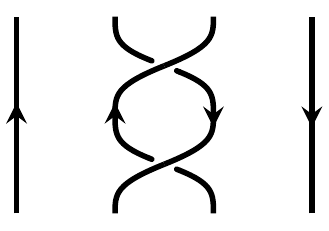}}.
    \end{align}        
    
    Let us find a condition, under which the elements of $G$ belong to vicinity (\ref{vicinity}) of $SU(d)$ group center, $d=r+1$. For this purpose, we represent $\bar T^2$ as
    \begin{equation}
        \bar T^2= \exp \left[i\varphi \diag\left(-\frac{1}{3}r(2r + 3N - 2)+2 m(m+N-1)\right)\right],  \qquad m = 0, \ldots r, \quad \varphi = 2\pi / (N + k) \label{Tbar2}
    \end{equation}
    and consider it's distance to the unit element $\id \in Z(SU(d))$
    \begin{equation}
        \|\bar T^2 - \id\|^2 = 4 \sum_{m=0}^{d-1} \sin^2\left[\frac{\varphi}2 \left(-\frac{1}{3}r(2 r+3 N-2)+2 m(m+N-1)\right)\right] < \frac12.
    \end{equation}
    Obviously, the element $\bar S \bar T^2 \bar S$ has the same distance to the unit element.
    Using the inequality $\sin x < x$, we obtain a simple sufficient condition for the inequality to hold
    \begin{equation}
        N+k > 
        \frac{2 \pi}{3} \sqrt{\frac{2}{5}r(r+1)(r+2)(16r^2+(30N-28)r+15N^2-30N+12)}. \label{r_cond}
    \end{equation}
    
    Now, let us show that the set (\ref{rgates}) satisfies the condition (\ref{ncond}), i.e. the space of matrices, which commutes with Lie-algebra elements in adjoint representation, corresponding to the elements of this set, is the space of matrices proportional to the unit matrix. For this purpose we first calculate these Lie-algebra elements. An element $x$, whose exponential gives $e^{x} = \bar T^2$, according to (\ref{Tbar2}) reads
    \begin{gather}
        x = i\varphi \diag\left(-\frac{1}{3}r(2r + 3N - 2)+2 m(m+N-1)\right) = i \diag (\varphi_1, \ldots \varphi_d), \label{x_def} \\ m = 0, \ldots r, \qquad \varphi \coloneqq 2\pi / (N + k), \nonumber 
    \end{gather}
    whereas exponentiation of $\bar S x\bar S$ gives the element $\bar S \bar T^2 \bar S$ due to $\bar S^2 = \id$. Thus, the set of Lie algebra elements, $X$, corresponding to (\ref{rgates}) reads
    \begin{equation}
        X = \{x, \; \bar S x\bar S\}. \label{rgates_log}
    \end{equation}
	Further, we first calculate the elements of $X$ in adjoint representation, and then prove that the condition (\ref{ncond}) is satisfied for the set above under some natural restriction on $k$.

	\subsection*{Calculation of $X$ in adjoint representation}    
    To find $\bar T^2$ and $\bar S \bar T^2 \bar S$ (equivalently, $x$ and $\bar S x \bar S$) in adjoint representation we use the specific properties of $\bar S$, namely
    \begin{equation}
        \bar S^2 = \id, \quad \bar S = \bar S^T.
    \end{equation}
    Therefore, $\bar S = e^{-\frac{i\pi}{2}(1-\bar S)}$ and $\bar S \, \bar T^2 \, \bar S = e^{-\frac{i\pi}{2}\bar S} \, \bar T^2 \,e^{\frac{i\pi}{2}\bar S}$.
    However, $i\bar S$ is not an element of $\mathfrak{su}_{\,d}$, $d=r+1$ Lie algebra (and $e^{\frac{i\pi}{2}\bar S}$ is not an element of $SU(d)$, correspondingly), since it's trace $\tr \bar S = -(r+1 \mod 2)$ can be non-vanishing. To overcome this difficulty we perform the following sequence of identical transformations 
    \begin{equation}
        \bar S \, \bar T^2 \, \bar S = e^{-\frac{i\pi}{2}\tilde S} \, \bar T^2 \,e^{\frac{i\pi}{2}\tilde S}, \qquad \tilde S =  \left\{\begin{aligned}
        &\bar S, & & \text{$r$ --- odd}, \\
        &\bar S + \frac1{r+1}\id, & & \text{$r$ --- even},
        \end{aligned}\right.
    \end{equation}
    where $i \tilde S$ does is an element of  $\mathfrak{su}_{\,d}$.
    Obviously, a similar formula takes place for $\bar S x \bar S$. Now, we can write down
    \begin{equation}
        \ad_{\bar S \, x \, \bar S} = 
        \Ad_{e^{-\frac{i\pi}{2}\tilde S}} \, \ad_{\,x} \,\Ad_{e^{\frac{i\pi}{2}\tilde S}}.
    \end{equation}
    Next, using the properties $\bar S^2 = \id$ and $\ad_{i\tilde S} = [i\tilde S, \bullet] = [i \bar S, \bullet]$, we get\footnote{In general, for any involutive matrix $ A $, that is, such that $ A^2 = \id$, the following identity takes place
    \begin{equation*}
        \Ad_{e^{\alpha A}} = e^{\alpha[A, \bullet]} = 1 + \frac14(\cosh2\alpha - 1) [A, [A, \bullet]] + \frac12 \sinh2\alpha [A, \bullet].
    \end{equation*}}
    \begin{equation}
        \mathcal O \coloneqq \Ad_{e^{\pm\frac{i\pi}{2}\tilde S}} = 1 + \frac12 [i\tilde S, [i\tilde S, \bullet]], \label{Ad_S}
    \end{equation}
    and, obviously, $\mathcal O^2 = \id$.
    
    To find an explicit form of elements (\ref{rgates_log}) in adjoint representation, one should choose the basis in $\mathfrak{su}_{\,d}$ algebra. Natural choice is the following 
    \begin{equation}
        X_{ij} = E_{ij} - E_{ji}, \quad Y_{ij} = i(E_{ij} + E_{ji}), \quad H_k = i\bigl(E_{kk} - \frac1d \sum_{l=1}^d E_{ll}\bigr), \qquad i, j, k = 1, \ldots d, \quad i<j,
    \end{equation}
    where the matrices $E_{ij}$ of size $d\times d$, defined as $(E_{ij})_{kl} = \delta_{il} \delta_{jk}$. This basis is useful because it allows to easily decompose an arbitrary element of $\mathfrak{su}_{\,d}$ as follows
    \begin{equation}
        A = \sum_{k<l} (A^{[kl]} X_{kl} - i A^{(kl)} Y_{kl}) - i \sum_{k=1}^d A^{kk} H_k, \qquad A \in \mathfrak{su}_{\,d},
    \end{equation}
    where $A^{[kl]}$ and $A^{(kl)}$ are antisymmetric and symmetric part of $A$, respectively. However, this basis is overcomplete, because $\sum_{k=1}^d H_k = 0$, and, in addition, is not orthonormal. To avoid corresponding difficulties, instead of $H_i$ we introduce the following elements
    \begin{equation}
        \tilde H_k = \left(\frac2{k(k+1)}\right)^{\!1/2} \left(\sum_{l=1}^k H_l - k H_{k+1}\right) = \sum_{l=1}^d Q^l{}_{k} H_l, \qquad k = 1, \ldots (d-1).
    \end{equation}
    Thus the set $\{X_{ij}, Y_{ij}, \tilde H_k\}$, where $i, j = 1, \ldots d$, $i<j$, and $k=1, \ldots (d-1)$, constitutes orthonormal basis in $\mathfrak{su}_{\,d}$ with respect to scalar product $(A, B) = -\frac12 \tr(A B)$. Elements $H_k$ can be expressed in terms of $\tilde H_k$ as follows
    \begin{equation}
        H_k = \frac12 \sum_{l=1}^{d-1} \tilde Q^l{}_{k} \tilde H_l, \qquad 
        \tilde Q \coloneqq Q^T.
    \end{equation}
    Using the explicit expressions for the commutators
    \begin{align}
        [X_{ij}, X_{kl}] &= X_{il} \delta_{jk} - X_{ik} \delta_{jl} - X_{jl} \delta_{ik} + X_{jk} \delta_{il}, \\
        [X_{ij}, Y_{kl}] &= 
        Y_{il} \delta_{jk} - Y_{jk} \delta_{il} + Y_{ik} \delta_{jl} - Y_{jl} \delta_{ik} + 2(H_j - H_i)(\delta_{ik}\delta_{jl} - \delta_{il} \delta_{jk}),  \label{XY_comm}\\
        [Y_{ij}, Y_{kl}] &= X_{kj} \delta_{il} + X_{ki} \delta_{jl} + X_{lj} \delta_{ik} + X_{li} \delta_{jk}, \\
        [H_i, H_j] &= 0, \\
        [X_{ij}, H_k] &= Y_{ik} \delta_{jk} - Y_{jk} \delta_{ik} , \\
        [Y_{ij}, H_k] &= X_{ki} \delta_{kj} + X_{kj} \delta_{ki},
    \end{align}
    we can write the expressions for $\ad_x$ and $\ad_{\bar S \, x \, \bar S} = \calO \ad_x \calO$ with $\calO$ defined in (\ref{Ad_S}). 
    
    We will represent elements of $\mathfrak{su}_d$ in adjoint representation as $3\times 3$ block-matrices of whole dimension $(d^2 - 1) \times (d^2 - 1)$, and blocks, corresponding to $X$, $Y$ or $\tilde H$ generators. Thus, for an arbitrary $A \in \mathfrak{su}_d$ we have
    \begin{equation}
    	\ad_A =
    	\begin{pmatrix}
    		A_{XX} & A_{XY} & A_{X\tilde H} \\
    		A_{YX} & A_{YY} & A_{Y\tilde H} \\
    		A_{\tilde HX} & A_{\tilde H Y} & A_{\tilde H \tilde H}
    	\end{pmatrix},
    \end{equation}
	where, for example $A_{XY}$, $A_{YY}$, and $A_{\tilde HY}$ are defined by
	\begin{equation}
		[A, Y_{ij}] = \sum_{k<l}\Bigl[(A_{XY})^{kl}{}_{ij} X_{kl} + (A_{YY})^{kl}{}_{ij} Y_{kl}\Bigr] + \sum_{k=1}^{d-1}(A_{\tilde HY})^{k}{}_{ij} \tilde H_{k}.
	\end{equation}
	Therefore, $XX$, $XY$, $YX$, $YY$ blocks are square $\frac{d(d-1)}2 \times \frac{d(d-1)}2$ matrices, and each row and column of those is numbered by the pair of indices like $ij$, where $i<j$. Similarly, $X \tilde H$, $Y \tilde H$ blocks are $\frac{d(d-1)}2 \times (d-1)$ matrices, $\tilde H X$, $\tilde H Y$ blocks are $(d-1) \times \frac{d(d-1)}2$ matrices, and $\tilde H \tilde H$ block is $(d-1)\times (d-1)$ matrix.
    
    Using the prescription above, and the decomposition
    \begin{equation}
        x = \sum_{i=1}^d \varphi_i H_i,
    \end{equation}
    where $\varphi_i$ are defined in (\ref{x_def}), we immediately obtain the following expression for $x$ in adjoint representation
    \begin{align}
        \ad_x = \begin{pmatrix}
            0 & D & 0\\ -D & 0 & 0 \\ 0 & 0 & 0
        \end{pmatrix}, \qquad D^{ij}{}_{kl} = \delta^{i}_k \delta^{j}_l (\varphi_l - \varphi_k),  \label{ad_x}
    \end{align}
    where $D$ is thought of as a diagonal matrix.
    
    Similarly, from the decomposition
    \begin{equation}
        i\tilde S = \sum_{i<j} \tilde S^{ij} Y_{ij} + \sum_{k=1}^{d} \tilde S^{kk} H_k, 
    \end{equation}
    we obtain an expression for $i\tilde S$ in adjoint representation
    \begin{equation}
    	\ad_{i\tilde S} = \begin{pmatrix}
    		0 &  M & N\\ M' & 0 & 0 \\ N' & 0 & 0
    	\end{pmatrix},
    \end{equation}
	where the block components read\footnote{Here, $\bar S$ appears instead of $\tilde S$, since $[\tilde S, \bullet] = [\bar S, \bullet]$.}
    \begin{align}
        M^{ij}{}_{kl} &= 2( \delta^{[i}_k \bar S^{j]}_l + \delta^{[i}_l \bar S^{j]}_k), &
        M'^{\,ij}{}_{kl} &= - M^{kl}{}_{ij} = - 2( \delta^i_{[k} \bar S^j_{l]} + \delta^j_{[k} \bar S^i_{l]}), \\ 
        N^{ij}{}_{k} & = \bar S^{ij}(Q^i{}_k - Q^j{}_k), &
        N'^{\,i}{}_{kl} &= -N^{kl}{}_{i} = -(\tilde Q^i{}_k - \tilde Q^i{}_l) \bar S_{kl}.
    \end{align}
    Now, using the expression (\ref{Ad_S}), we can write an expression for $\Ad_{e^{\frac{i\pi}2\tilde S}}$ as
    \begin{equation}
        \calO = \Ad_{e^{\frac{i\pi}2\tilde S}} = \id + \frac12 (\ad_{i\tilde S})^2 = \id + \frac12\begin{pmatrix}
           M M'+ N N' &  0 & 0\\ 0 & M'M & M'N \\ 0 & N' M &  N' N
        \end{pmatrix} =: 
        \begin{pmatrix}
            S_1 &  0 & 0\\ 0 & S_2 & R \\ 0 & R^T &  S_3
        \end{pmatrix}, \label{Ad_expS}
    \end{equation}
    where the block components are
    \begin{gather}
        (S_1)^{ij}{}_{kl} = 2 \, \bar S^{[i}_k \bar S^{j]}_l, \qquad (S_2)^{ij}{}_{kl} = 2 \, \bar S^{(i}_k \bar S^{j)}_l, \qquad (S_3)^i{}_j = \frac12 \sum_{k,l=1}^d  \tilde Q^i{}_k \, (\bar S^k_l)^2 \, Q^l{}_j,\\
         R^{ij}{}_k = 2 \sum_{l=1}^d \bar S^i_l \bar S^j_l\, Q^l{}_k, \qquad (R^T)^i{}_{kl} = 2 \sum_{j=1}^d \tilde Q^i{}_j \, \bar S^j_k \bar S^j_l. \label{R_block}
    \end{gather}
    Thus, $\ad_{\bar S \, x \, \bar S} = \calO \ad_x \calO$ can be obtained just by the multiplication of (\ref{ad_x}) and (\ref{Ad_expS}). From $\calO^2 = \id$ one gets the following identities relating $S_1$, $S_2$, $S_3$ and $R$
    \begin{equation}
        (S_1)^2 = \id, \qquad (S_2)^2 + R R^T = \id, \qquad (S_3)^2 + R^T R = \id, \qquad S_2 R + R S_3 = 0. \label{Ad_S_cons}
    \end{equation}

    \subsection*{Proof of (\ref{ncond}) for the set (\ref{rgates_log})}
    Now, we are prepared for checking the property (\ref{ncond}) of the set (\ref{rgates_log}). First of all, if $\varphi_k - \varphi_l \ne 0$ for $k\ne l$, the space $C(\ad_x) = \{L \in \mathfrak{gl}_{d^2-1} : [L, \ad_x] = 0\}$ of matrices commuting with $\ad_x$, consist of elements of the form
    \begin{equation}
        L = \begin{pmatrix}
            D_1 & D_2 & 0 \\ 
            -D_2 & D_1 & 0 \\
            0 & 0 & K
        \end{pmatrix}, \label{C_ad_elem}
    \end{equation}
    where $D_1$, $D_2$ are arbitrary diagonal matrices of the size $\frac{d(d-1)}2 \times \frac{d(d-1)}2$, whereas $K$ is an arbitrary matrix of the size $(d-1) \times (d-1)$. Thus, to check (\ref{ncond}), we should show that the only matrices $L \in C(\ad_x)$ commuting with $\ad_{\bar S \, x \, \bar S}$ are proportional to identity matrix. 
    
    Using the following equality
    \begin{equation}
        [L, \ad_{\bar S \, x \, \bar S}] = \calO [\calO L \calO, \ad_x] \calO, \qquad \calO = \calO^T = \Ad_{e^{\pm\frac{i\pi}{2}\tilde S}},
    \end{equation}
    we conclude that $C(\ad_{\bar S \, x \, \bar S}) = \calO C(\ad_x) \calO$, so now we should prove that $C(\ad_x) \cap \calO C(\ad_x) \calO = \{\lambda \id, \lambda\in \mathbb R\}$, i.e. the matrix $L \in C(\ad_x)$, such that $\calO L \calO$ is also in $C(\ad_x)$, must be proportional to the identity matrix.
    
    Since symmetric and antisymmetric parts of $L = L_s + L_a$ transform under conjugation by orthogonal matrices irreducibly, we can consider them separately. Let us begin with the antisymmetric part $L_a$
    \begin{gather}
        \calO L_a \calO = 
        \begin{pmatrix}
            S_1 & 0 & 0 \\
            0 & S_2 & R \\
            0 & R^T &  S_3
        \end{pmatrix}
        \begin{pmatrix}
             0 & D_2 & 0 \\
             -D_2 & 0 & 0 \\
             0 & 0 & A
        \end{pmatrix}
        \begin{pmatrix}
            S_1 & 0 & 0 \\
            0 & S_2 & R \\ 
            0 & R^T & S_3
        \end{pmatrix} \nonumber \\ = 
        \begin{pmatrix}
            0 & S_1 D_2 S_2 & S_1 D_2 R^T \\ 
            -S_2 D_2 S_1 & R A R^T & R^T A S_3 \\ 
            -R^T D_2 S_1 & S_3 A R^T &  S_3 A S_3
        \end{pmatrix}.
    \end{gather}
    Comparing with (\ref{C_ad_elem}), we conclude that $D_2$ should vanish, because $S_1 D_2 S_2$ is not symmetric and, therefore, is not diagonal. We also conclude that if $R$ is of maximal rank, then $A$ should vanish because the block $R A R^T$ should vanish. Similarly, for the symmetric part $L_s$ we have
    \begin{align}
        \calO L_s \calO &= 
        \begin{pmatrix}
            S_1 & 0 & 0 \\
            0 & S_2 & R \\
            0 & R^T & S_3
        \end{pmatrix}
        \begin{pmatrix}
             D_1 & 0 & 0 \\ 
             0 & D_1 & 0 \\
             0 & 0 & \Sigma
        \end{pmatrix}
        \begin{pmatrix}
            S_1 & 0 & 0 \\
            0 & S_2 & R \\
            0 & R^T & S_3
        \end{pmatrix} \nonumber\\ &= 
        \begin{pmatrix}
            S_1 D_1 S_1 & 0 & 0 \\ 
            0 & S_2 D_1 S_2 + R \Sigma R^T & S_2 D_1 R + R \Sigma S_3 \\
            0 & R^T D_1 S_2 + S_3 \Sigma R^T & R^T D_1 R + S_3 \Sigma S_3
        \end{pmatrix}.
    \end{align}
    Suppose $S_1$ is such that $S_1 D_1 S_1$ is diagonal only for $D_1 = d_1 \id$. If so, due to (\ref{Ad_S_cons}) we get $S_1 D_1 S_1 = d_1 \id$. Therefore, central block should be equal to $d_1 \id$, that leads to equality $R (\Sigma - d_1 \id) R^T = 0$. If $R$ is of maximal rank, brackets should vanish and $\Sigma = d_1 \id$. Finally, we get $\calO L\calO = d_1 \id$ as required, if our assumptions about diagonal property of $D_1$, and rank maximality of $R$ are valid.
    
    Thus, we still should prove that $S_1 D_1 S_1$ is diagonal only for $D_1 = d_1 \id$ and that the matrix $R$ has the maximal rank, under some conditions.
    
    If all diagonal elements of $D_1$ are different, the only matrix $S_1$, which gives diagonal $S_1 D_1 S_1$, is permutation matrix. Obviously it is not the case. Now, suppose $D_1$ has $M$ coinciding diagonal elements. Then, there is a permutation matrix $\sigma$ such that $\sigma^T D_1 \sigma$ has the following block-diagonal form
    \begin{equation}
        \sigma^T D_1 \sigma = 
        \begin{pmatrix}
            d_1 \id & 0 \\
            0 & \tilde D_1
        \end{pmatrix},
    \end{equation}
    where $\tilde D_1$ is diagonal matrix, whose diagonal elements are pairwise different.
    Then, we rewrite $S_1 D_1 S_1 = (S_1 \sigma )\sigma^T D_1 \sigma (\sigma^T S_1)$, so we can deduce the most general form of $S_1$ that gives diagonal matrix $S_1 D_1 S_1$
    \begin{equation}
        S_1 = \sigma \begin{pmatrix}
            \tilde S_1 & 0 \\
            0 & \id
        \end{pmatrix} \tilde \sigma,
    \end{equation}
    where $\tilde S_1$ is an arbitrary orthogonal $M \times M$ matrix, while $\tilde \sigma$ is an arbitrary permutation matrix. Thus, we conclude that $S_1$ can have at most $M$ nonvanishing elements at each row and column. Let us find out when it's not the case. Consider particular row of $(S_1){}^{ij}{}_{kl}=2\tilde S^{[i}_k \tilde S^{j]}_l$, namely $k=1$, $l=2$. Using the explicit expressions
    \begin{align}
        \bar S_{1i} &= (-)^r \sqrt{[N-1][N+2i-3]} \frac{[N+i-3]![r]!}{[N+r-1]![i-1]!}, \label{S_0i} \\
        \bar S_{2i} &= (-)^{r+1} \sqrt{[N+1][N+2i-3]}
        \frac{[N+i-3]![r-1]!}{[N+r]![i-1]!}([i-1]^2-[N-1][r-i+1][N+r+i-1]),
    \end{align}
    we find that $(S_1){}^{ij}{}_{01} = h(i, j) (f(i) - f(j))$, where $h(i, j)$ is some multiple of $q$-numbers that don't vanish under condition
    \begin{equation}
        k > N + 4r - 2, \label{q_nonv}
    \end{equation}
    while $f(i) = [i-1]^2+[N-1][N+i-1][i-1]$ is monotonic function of $i$ under the same condition (\ref{q_nonv}), so that $f(i) - f(j) \ne 0$ for $i < j$. Therefore, we found the raw of $S_1$ with all non-vanishing elements and conclude that the number $M$ of coinciding elements of $D_1$ is maximal, i.e. $D_1 = d_1 \id$.
    
    To prove that the matrix $R$ has the maximal rank under some conditions, we choose its submatrix $\tilde R$ of maximal size and prove that it's not singular. Namely, we fix $i=1$ in (\ref{R_block}) to obtain
    \begin{gather}
     \tilde R^j{}{}_k \coloneqq R^{1j}{}_k = 2 (\bar S \tilde D Q)^j{}_k, \qquad \tilde D \coloneqq \diag (\bar S_{11}, \bar S_{12}, \ldots \bar S_{1d}), \\ 
     j=2, \ldots d, \qquad k = 1, \ldots (d-1). \nonumber
    \end{gather}
    Here $\bar S$, $\tilde D$, $Q$ are of size $(d-1)\times d$, $d\times d$ and $d\times (d-1)$, respectively. If all of them are of maximal rank, then $R$ is also of maximal rank as needed. While $Q$ and $\tilde S$ are of maximal rank as is, maximality of rank of $\tilde D$ implies $\bar S_{1i} \ne 0$, $i = 1, \ldots d$. Using the explicit expression (\ref{S_0i}), we found that none of $q$-numbers on the r.h.s. are zero if (\ref{q_nonv}) is satisfied. It's a sufficient condition of rank maximality of $R$. This completes the proof of universality of the set of gates (\ref{rgates}).

    \section{Discussion} \label{Disc_sec}
    
    In this paper we examined the universality of the Chern-Simons topological quantum computer --- a device that can be effectively described by the Chern-Simons topological quantum field theory with $SU(N)$ gauge group and the level $k$. For this purpose, we compared the evaluation of the Wilson loop to quantum algorithms, namely, preparation of a state, its unitary evolution, and subsequent measurement. In the case when the Wilson loop corresponds to a four-plat knot and symmetric representation $[r]$, evolution is determined by $\calR$-matrices acting on a Hilbert space of dimension $r+1$. We found that they are unitary under the condition
    \begin{equation}
    	k > N + 4r - 2.
    \end{equation}
    Further, we examined the universality of the set of $\calR$-matrices, and answered the question when an arbitrary unitary matrix can be represented as a product of $\calR$-matrices. Namely, using the recently proposed method \cite{qudit}, we found that the following condition is a sufficient universality condition
    \begin{equation}
    	k > 
    	\frac{2 \pi}{3} \sqrt{\frac{2}{5}r(r+1)(r+2)(16r^2+(30N-28)r+15N^2-30N+12)} - N = O(r^{5/2}). \label{r_cond_disc}
    \end{equation}
    Thus, in order to simulate $n$ qubits, we must have $r = 2^n - 1$ and, according to the inequality above $k = O(2^{5n/2})$. Thus, to perform efficient quantum computations by the way considered in the present paper, one should have the Chern-Simons topological quantum computer with sufficiently large (and, in fact, very large) values of level $k$.

	Demand of large values of $k$ as universality condition is rather natural from the following point of view. Consider the Solovay-Kitaev theorem \cite{kitaev,nielsen} that basically says that if one is able to approximate arbitrary unitary matrices with some zero-order accuracy (depending on its dimension), one also can approximate them with any desired accuracy. Moreover, the theorem provides a constructive algorithm for such an approximation \cite{nielsen}. We intend to demonstrate the universality with the use of this theorem and find that it also requires the large values of $k$.
	
	Let us first note that for large $k$ the gates $\bar T^2$, $\bar S \bar T^2 \bar S$, used in Sect. \ref{Univ_sec}, are close to the identity matrix. Next, let us represent an arbitrary unitary $\mathcal U$ as an element of $n$-parametric family of unitary matrices
	\begin{equation}
		\mathcal V(\alpha_i) = \bar T^{2\alpha_1} \bar S \bar T^{2\alpha_2} \bar S \ldots \bar T^{2\alpha_{n-1}} \bar S \bar T^{2\alpha_n} \bar S
	\end{equation}
	composed of the gates $\bar T^2$, $\bar S \bar T^2 \bar S$, in arbitrary, not necessary integer, powers. If $n \gtrsim d^2 - 1 = \dim SU(d)$ is sufficiently large and logarithms of these gates generate $\mathfrak{su}_{\,d}$ Lie algebra\footnote{According to Corollary 3.3 of \cite{qudit}, the proof of this fact coincides with the proof of the property (\ref{ncond}) provided in Sect. \ref{Univ_sec} of the present article.}, it is always possible (see Lemma 6.2 in \cite{control}). Then, we round $\alpha_i$ to the nearest integers $[\alpha_i]$, so $\mathcal V([\alpha_i])$ can be realized as a sequence of gates $\bar T^2$, $\bar S \bar T^2 \bar S$. Thus, $\mathcal V([\alpha_i])$ approximates $\mathcal U$ with the following accuracy
	\begin{equation}
		\|\mathcal V([\alpha_i]) - \mathcal U\| \le n \, \|\bar T^2 - \id \| = O(d^{9/2}/k), \label{appr_est}
	\end{equation}
	where we used (\ref{r_cond}). For $\mathcal V([\alpha_i])$ to be a valid zeroth-order approximation in the Solovay-Kitaev theorem, the following condition should be satisfied \cite{nielsen}
	\begin{equation}
		\|\mathcal V([\alpha_i]) - \mathcal U\| \le \left( 4 d^{3/4} \Bigl(\frac{d-1}2\Bigr)^{3/2} + 8 d^{1/4} \Bigl(\frac{d-1}2\Bigr)^{1/2}\right)^{\!\!-2} = O(d^{-9/2}).
	\end{equation}
	Comparing with (\ref{appr_est}), one can find that for $k = O(d^{9})$, an arbitrary unitary $\mathcal U$ can be approximated with the use of Solovay-Kitaev theorem. This condition is more restrictive than (\ref{r_cond_disc}) but gives an intuitive answer to the question, why the universality does hold for sufficiently large $k$, and, in addition, it provides a constructive algorithm of unitary matrix approximation, when the condition $k = O(d^{9})$ is satisfied.

	\section*{Acknowledgments}
	We are grateful for very useful discussions to V. Alekseev, A.
	Popolitov, A. Sleptsov, and N. Tselousov.
	This work was supported by Russian Science Foundation grant No 18-71-10073.

\end{document}